\documentclass{kluwer}    
\usepackage[dvips]{graphicx}
\newdisplay{guess}{Conjecture}

\begin{document}                                                                                   
\begin{article}
\begin{opening}         
\title{Observational Properties of Jets \\
in Active Galactic Nuclei} 
\author{Gabriele \surname{Giovannini}}  
\runningauthor{Gabriele Giovannini}
\runningtitle{Observational properties of jets in AGN}
\institute{Dipartimento di Astronomia dell'Universita' di Bologna and \\
Istituto di Radioastronomia del CNR}
\date{December 15, 2003}

\begin{abstract}
Parsec scale jet properties are shortly presented and discussed.
Observational data are used to derive constraints on the jet
velocity and orientation, the presence of velocity structures, and  the 
connection between the pc and kpc scale.
Two peculiar sources with limb-brightened jets: 1144+35 
and Mkn 501 are discussed in detail.
\end{abstract}

\end{opening}           

\section{Introduction}  

Jets are present in many galactic and extragalactic objects and may have
different origin and properties. I will focus here to extragalactic radio jets
from the parsec (pc) to the kiloparsec (kpc) scale.
These jets are among the largest coherent fluid flow patterns in 
the universe.
They can have very different working surface (from pc to Mpc), and 
they can grow of
a factor $\sim$ 10$^6$. Moreover the strength of the magnetic field in 
the jet and the particle density can change of a large factor ($\sim$ 10$^4$).

To study jet properties, their origin and evolution we need to start from the
 pc scale region and to follow their properties up to the Mpc 
scale. At present only Very Long Baseline Interferometry (VLBI) observations 
in the radio band can have the angular
resolution to obtain pc scale images in extragalactic radio sources.
With these technique radio telescopes across continents are combined to form a
global virtual radio telescope with a size of the earth which provides
the highest resolution achievable in astronomy.

Thanks to VLBI observations I will present and discuss here observational 
properties of pc scale jets and I will compare these results with kpc scale 
jet properties in low and high power radio sources.

\section{Jet Morphology}

To get new insight in the study of radio jets at pc resolution,
it is important to select source samples from low frequency catalogues,
where the source
properties are dominated by the unbeamed extended emission
and are not affected by observational biases related to orientation effects.
To this aim,
we undertook a project of observations of a complete sample of radio
galaxies selected from the B2 and 3CR catalogs 
with z $<$ 0.1 (i.e. no constrain
on the core flux density): the {\it Complete Bologna Sample} (CBS; Giovannini
et al. in preparation).
This sample consists of 95 sources. At present 53 on 95 sources have
been studied with VLBI observations. I will use published results and these 
preliminary data to
discuss the jet properties on the pc scale.
 
Many of the extended FR I and FR II sources have a one-sided structure on the
pc scale (18 (34\%) FR I and 15 (28\%) FR II). From these high resolution 
images, we note that high and low power sources appear very similar.
Parsec scale images are very similar and it is not possible to discriminate
FR I from FR II sources using VLBI images.
In high and low power sources with
relatively faint radio cores, the number of two-sided sources increases
as expected from unified scheme models. 

\begin{figure}[H]
\tabcapfont
\centerline{%
\begin{tabular}{c@{\hspace{6pc}}c}
\includegraphics[width=2in]{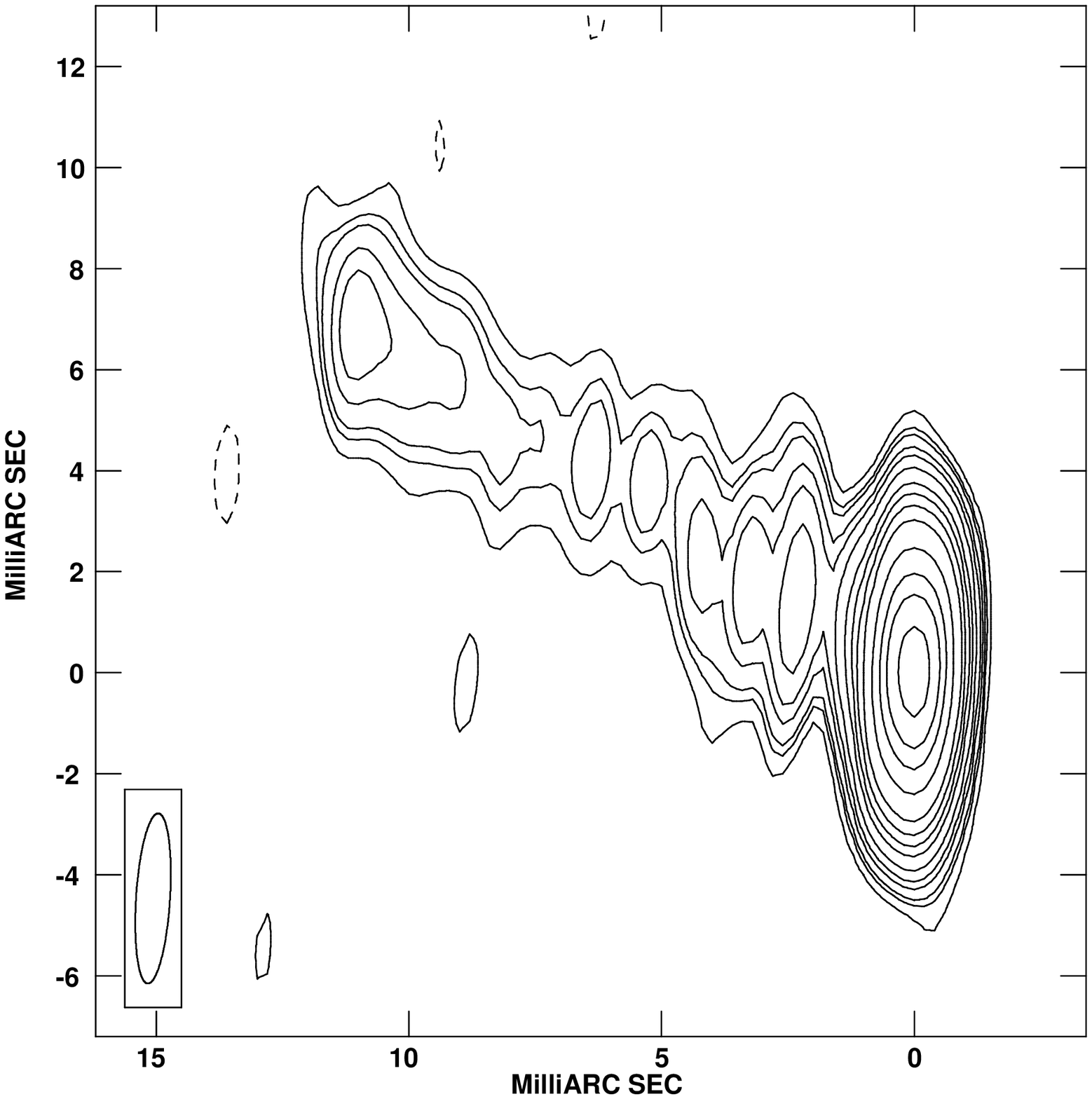} &
\includegraphics[width=2in]{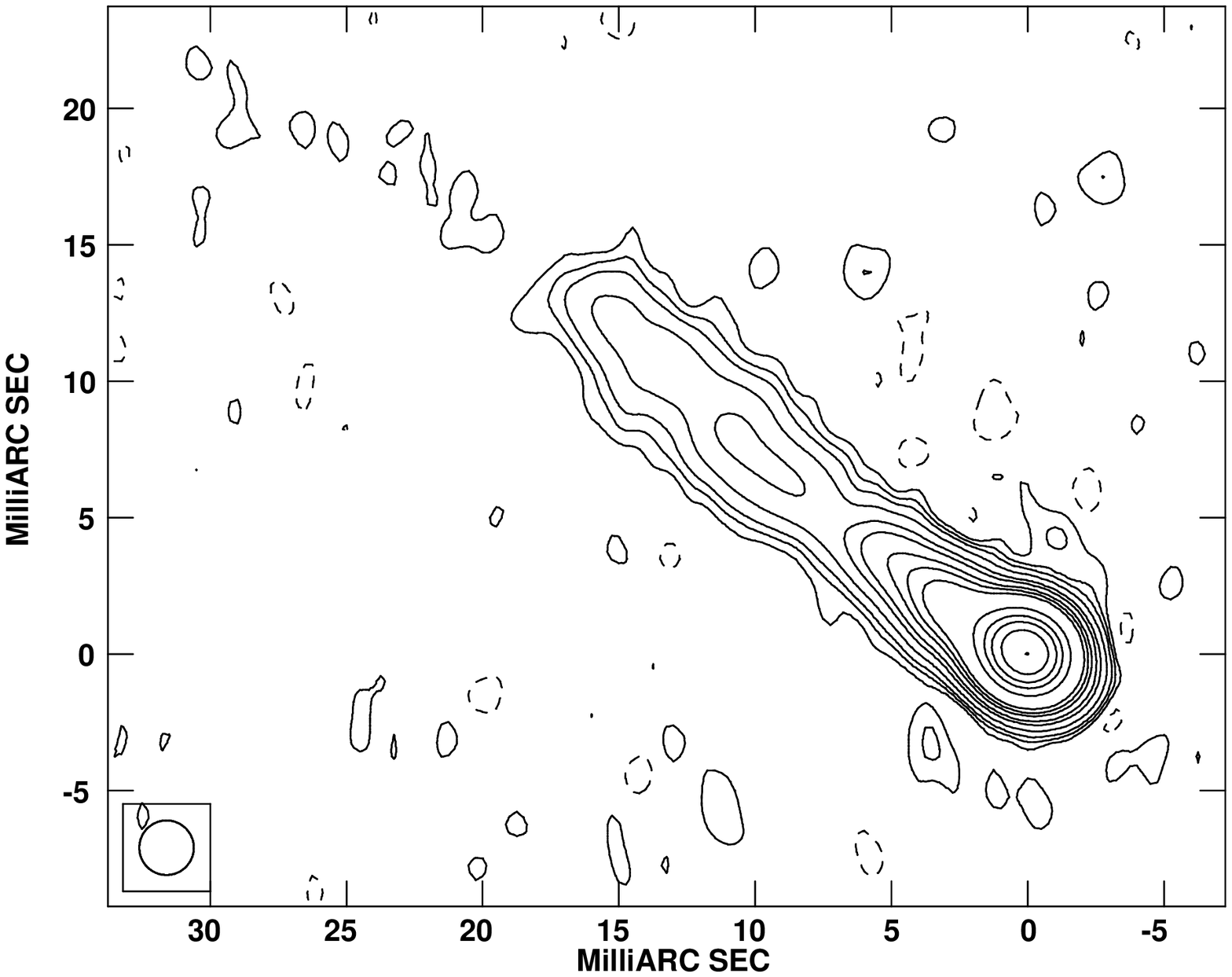} \\
a.~~ 3C 382 & b.~~ 3C 66B
\end{tabular}}
\caption{VLBI image of 3C382, a Broad Line FR II RG and 3C 66B a 
FR I RG.}
\label{fig1}
\end{figure}

Among the observed sources from
our sample there are
7 FR I and 4
FR II galaxies with two-sided jets (21\%). In contrast, there
are 7/65 (11\%) symmetric sources in the Pearson-Readhead sample 
(Pearson \& Readhead 1988) and 18/411 (4.4\%) in the combined
PR and CJ samples (Taylor et. al 1994; Polatidis et al. 1995).
 
We note that all two-sided FR II radio
galaxies are Narrow Line (NL) objects confirming that Broad Line Radio 
Galaxies (BLRG) are
oriented at least as close to the line-of-sight as quasars.

\begin{figure}
\centerline{
\includegraphics[width=24pc]{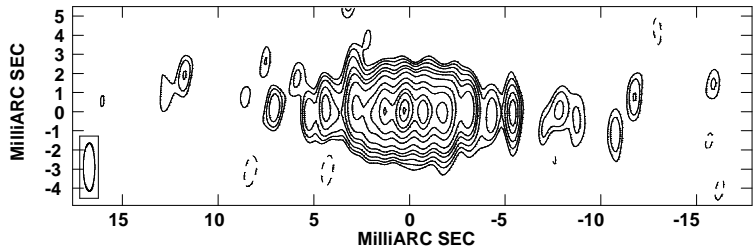}}
\caption{VLBI image at 5 GHz of 3C 452 , a Narrow Line FR II radio galaxy}
\label{fig2}
\end{figure}

\begin{figure}
\centerline{
\includegraphics[width=24pc]{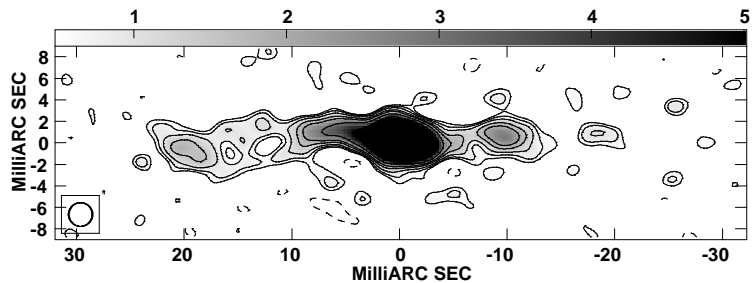}}
\caption{VLBI image at 5 GHz of 3C 338 , a FR I radio galaxy}
\label{fig3}
\end{figure}

In most sources we find a good agreement between
the pc and kpc scale structures. The comparison between the VLA and VLBA
jet Position Angle (PA) shows only in one source a
large difference in the jet orientation ($\sim$ 90$^\circ$) and in
2 other sources a difference of 50 and 30 degrees.
This result support the model where the large
distortions detected in BL-Lac sources and quasars are due to small bendings 
amplified
by the small angle of orientation of these objects.

We have also compared the correlated flux in the shortest VLBA baselines
with the core arcsecond flux density. In 6 sources this comparison
was not
possible because of the too large variability or to the presence of compact
steep spectrum structures, among the other 47 sources, 33 (70\%) have a
correlated flux larger than 70\% of the arcsecond core flux density.
Therefore in these sources we have mapped most of the small scale structure
and we are able to properly connect the pc to the kpc structure. In the
reamining 14 (30\%) sources we are missing in the VLBA images a relevant
fraction of the arcsecond core flux density (larger than 90\% in a few cases).
This result suggests the presence of relevant sub-arcsecond structures not
visible in VLBA images because of the lack of short baselines and unresolved
in VLA images. 

\section{Jet Velocity}

\subsection{Proper Motion}

Many AGNs contain compact radio sources with different components which
appear to move apart. Multi epoch studies of these sources allow a direct
measure of the apparent jet pattern velocity ($\beta_a$c). The observed
distribution of the apparent velocity shows a large range of values
(Vermeulen and Cohen, 1994; Kellerman et al., 2000).
From the measure of $\beta_a$ we can
derive constraints on $\beta_p$ and $\theta$ where $\beta_p$c is the intrinsic
velocity of the pattern flow and $\theta$ is the jet orientation with respect
to the line of sight:

\begin{equation} 
\beta_p = \beta_a/(\beta_a cos\theta + sin\theta)
\end{equation}

A main problem is to understand the difference between the bulk and pattern
velocity. In few cases where proper motion is well defined and the bulk
velocity is strongly constrained, there is a general agreement between the
pattern velocity and the bulk velocity (see e.g. NGC 315 in Cotton et al., 
1999,
and 1144+35, here). However, in the same source we can have different
pattern velocities as well as stationary and high velocity moving structures.
Moreover, we note that in many well
studied sources the jet shows a smooth and uniform surface brightness and
no (or very small) proper motion (as in the case of Mkn 501,
Giroletti et al. 2004, and M87 in the region at $\sim$ 1 pc from the
core, Junor et al., 1999).

\subsection {Bulk Velocity}

Assuming that the jets are intrinsically symmetric we can use relativistic
effects to constrain the jet bulk velocity $\beta$c and orientation with
respect to the line of sight ($\theta$) as following:

\begin{itemize}

\item
jet - counterjet ratio

\item
core dominance

\item
synchrotron Self Compton emission

\item
arm length ratio

\item
brightness temperature

\end{itemize}

I will discuss here only the first two points which are the most used in 
our sample and literature.

\subsubsection {Jet - Counterjet ratio}

Assuming that the jets are intrinsically symmetric we can use the observed
jet to counter-jet brightness ratio R to constrain the jet
bulk velocity $\beta$c
and its orientation with respect to the line of sight ($\theta$):

\begin{equation} 
R = (1+\beta cos\theta)^{2+\alpha} (1-\beta cos\theta)^{-(2+\alpha)}
\end{equation}

where $\alpha$ is the jet spectral index (S($\nu$) $\propto$ $\nu^{-\alpha}$).
Some problems can be related to this measurement:
free free absorption may affect the observed jet brightness, moreover,
we cannot derive strong constraints in intrinsically low luminosity jets:
e.g. in 3C264 where a well studied optical and radio jet is present
(Lara et al., 1999) the highest j/cj ratio is R $>$ 37 which implies
that the source is
oriented at $\theta$ $<$ 52$^\circ$ with a jet moving with $\beta$ $>$ 0.62.

\subsubsection{Core Dominance}

The core radio emission measured at 5 GHz, at arcsecond resolution is dominated
by the Doppler-boosted pc-scale relativistic jet.
The source radio power measured at low frequency (e.g. 408 MHz), instead,
is due
to the extended emission, which is not affected by Doppler
boosting. At low frequency the observed core radio emission is
not relevant since it is mostly self-absorbed.
Given the existence of a general correlation between the core and total
radio power discussed in  Giovannini et al., 2001,
we can
derive the expected intrinsic core radio power from the {\it unboosted}
total radio power using the estimated best fit correlation (continuum line in 
Fig. 4):

\begin{equation}
log P_c = (0.62\pm0.04) log P_t + (7.6 \pm 1.1)
\end{equation}

The comparison between the expected intrinsic core radio
power and the observed core radio power will give constraints on the jet
velocity and orientation (Giovannini et al., 2001).
We note that the core radio power is best measured at 5 GHz where
it is dominant because of the steep spectrum of the extended emission,
self-absorption is not relevant, and high angular
resolution images allow us to separate the core from the extended jet
emission.\\
The large dispersion in the core
radio power visible in Fig. 4 is expected because
of the strong dependance of the observed core radio power on $\theta$ and
$\beta$.

\begin{figure}
\centerline{
\includegraphics[width=24pc]{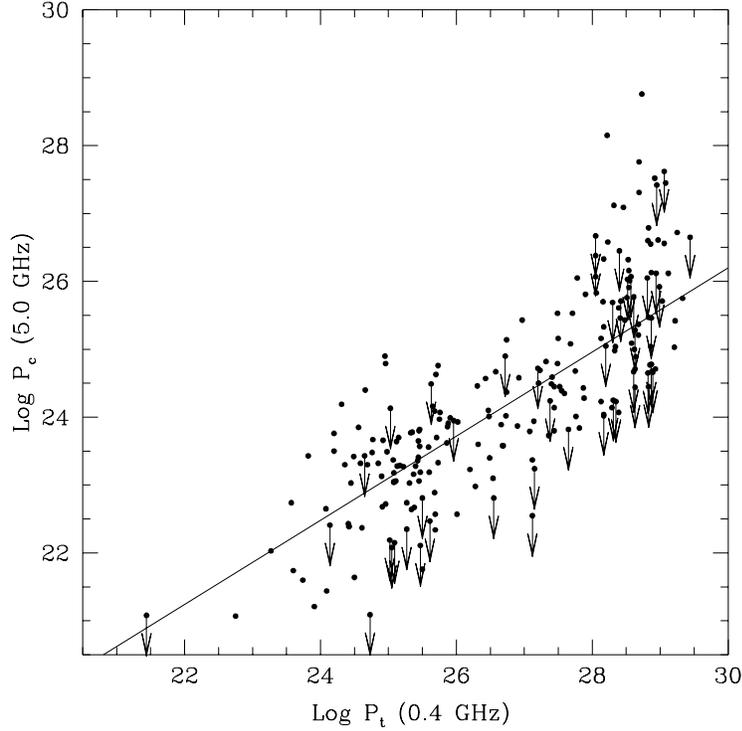}}
\caption{Arcsecond core radio power at 5 GHz (P$_c$) versus total radio power
at 0.4 GHz (P$_t$). Arrows are upper limits when a nuclear emission was
not detected. The line is from the equation in Sect. 3.2.2}
\label{fig4}
\end{figure}

From the data dispersion, assuming that no selection effect is present in
the source orientation ($\theta$ = 0$^\circ$ to 90$^\circ$),
we can derive that the
jet Lorentz factor $\Gamma$
has to be $<$ 10 otherwise we should observe a larger core radio power
dispersion.

\subsection{Results}

To derive statistical properties of radio jets on the pc scale, we
used all observational data for the 51 sources in our sample with VLBI data.
We found that in all sources pc scale jets move at high velocity. No
correlation has been found between the jet velocity and  the core or total
radio power.
Highly relativistic parsec scale jets
are present regardless of the radio source power. Sources with a different kpc
scale morphology, and
total radio power have pc scale jets moving at similar velocities.

We used the estimated $\beta$ and $\theta$ to derive
the Doppler factor $\delta$ for each source,
and the corresponding intrinsic core radio power (assuming $\alpha$ = 0):

\begin{equation} 
P_{c-observed} = P_{c-intrinsic} \times \delta^2
\end{equation} 
  
We found a good correlation between  P$_{c-intrinsic}$ and P$_{t}$ with a
small dispersion
since plotting P$_{c-intrinsic}$, we removed the spread due to
the different orientation angles (Fig. 5).
We found that a Lorentz factor $\Gamma$ in the
range 3 to 10 is consistent with the observational data.
The Lorentz factor cannot be $>$ 10 for the previous considerations
(Sect. 3.2).
It cannot be $<$ 3 to remove the dispersion due to the different
source orientations.
The result that sources with different P$_{t}$ and kpc scale morphology
show the same correlation, implies that {\bf all pc scale
jets have
a similar velocity}.
Two sources do not follow the general correlation: M87 where
we should
have a higher jet velocity to fit with the general correlation and
3C 192 which core radio power is lower than expected (it could be in a
pre-relic phase).

\begin{figure}[H]
\tabcapfont
\centerline{%
\begin{tabular}{c@{\hspace{0.5pc}}c}
\includegraphics[width=3.0in]{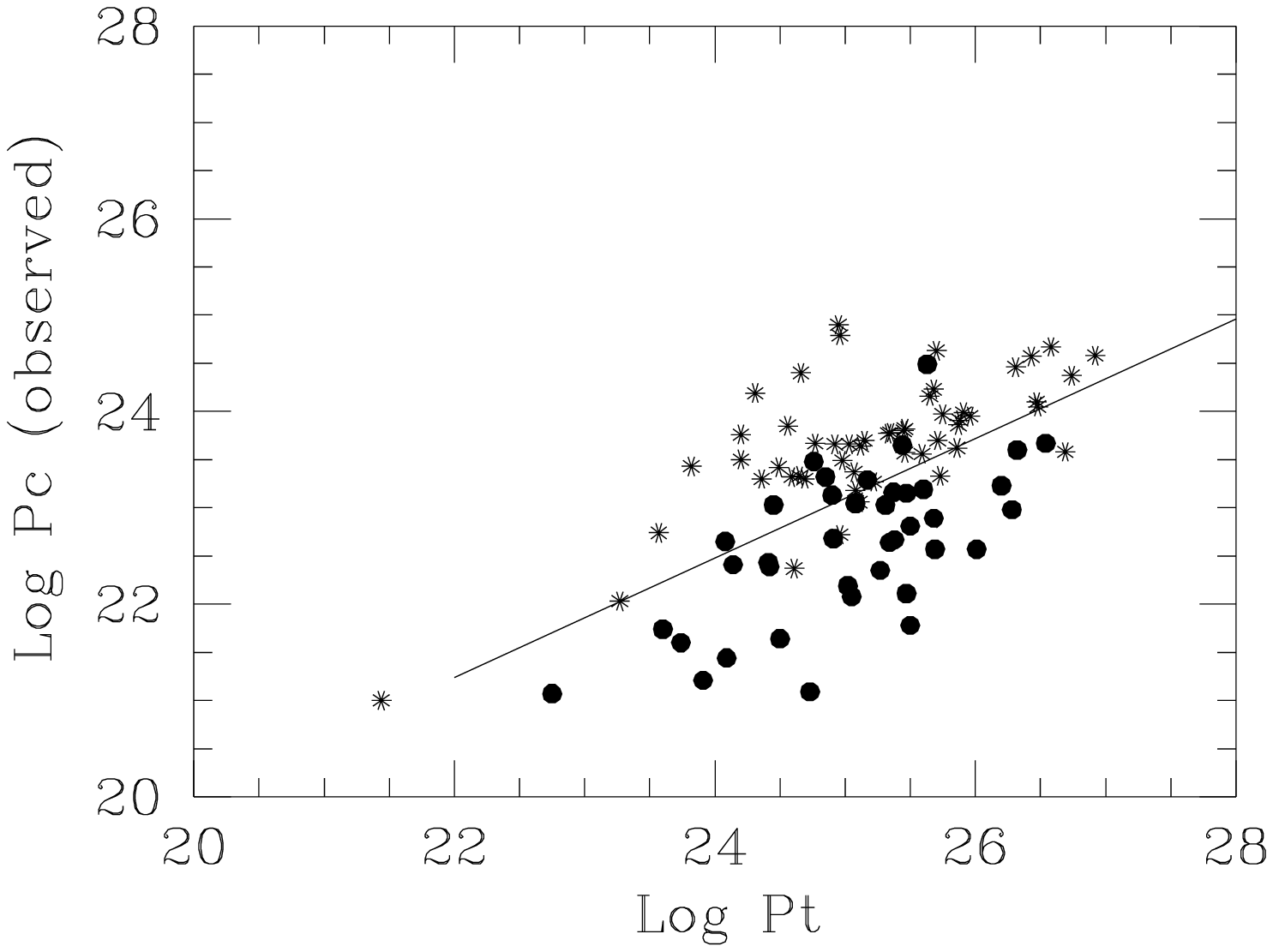} &
\includegraphics[width=3.0in]{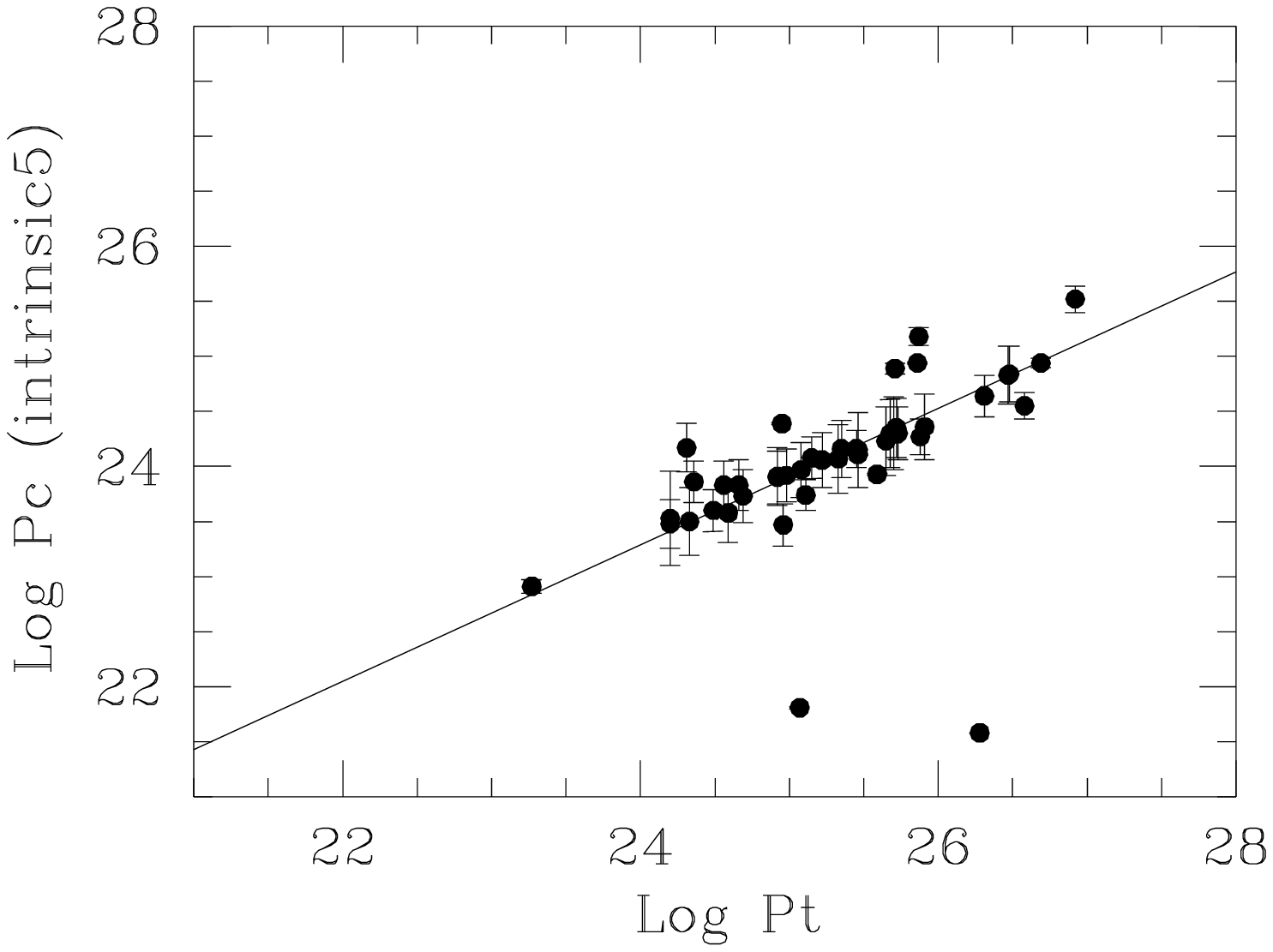} \\
a.~~ observed data & b.~~  intrinsic data.
\end{tabular}}
\caption{Observed core radio power versus total radio power (a).
Intrinsic core radio power (see Sect. 3.3) versus total radio power (b)}
\label{fig5}
\end{figure}

\section{Acceleration and Deceleration in Jets}
 
In some sources evidences of increasing velocity in pc scale jets
have been found: e.g. M87 (Biretta et al., 1995);
3C84 (Dhawan et al., 1998); Cygnus A (Krichbaum et al., 1998).
However, in these sources the jet velocity was measured from the jet
apparent motion. Thus the increasing velocity could be
non intrinsic but due to a change in the jet direction
or to a change in the jet pattern velocity unrelated
to the jet bulk velocity. In NGC 315 Cotton et al. (1999) found
an increasing
jet velocity in the 5 inner pc from the core both from proper motion
measurements AND
from the sidedness ratio. Recently an increasing bulk velocity has been
found also in NGC 6251 by Sudou et al. (2000). 

In many low power sources there are
observational evidences (Laing et al. 2003) that jets are relativistics at 
their beginning and
strongly decelerate within a few kpc from the core (see e.g. 3C449,
Feretti et al., 1999).\\
In FR II radio sources kpc scale jets are still affected by Doppler
boosting effects, but the jet sidedness ratio decreases with the
core distance
implying a velocity decrease. Observational data are consistent with $\Gamma$
$\sim$ 2 on the kpc scale (Bridle et al., 1994).

The discovery of X-Ray emission from jets on the kpc scale is an evidence in
favor of relativistic beaming even on large scales (e.g. Sambruna, 2003).

\section{Velocity Structures}
 
An evident limb-brightened jet morphology on the pc scale is present in some
FR I sources as 1144+35, Mkn 501, 3C264, M87, 0331+39 (Giovannini et al., 
2001),
and at least in one high power radio source: 1055+018 (Attridge et al., 1999). 
We interpret the
limb-brightened structure as due to a different Doppler boosting effect in
a two-velocity relativistic jet. If the source is oriented at a relatively
large angle with respect to the line of sight, the inner very high
velocity spine
could be strongly deboosted, while the slower external layer could be boosted
and appear brighter than the inner jet region. Therefore only sources in
a small range of $\theta$ will appear limb-brightened. For this reason, and
for the observational
difficulties of transversally resolving the radio jets,
we expect that the number of sources exhibiting limb-brightened jets is low,
as observed.
 
At present it is not clear if the velocity structure is strictly related
to the jet interaction with the ISM as suggested by Giovannini et al. (2001)
or it is an intrinsic jet property. The presence of this structure very near to
the central core as in M87 (Junor et al., 1999) and Mkn 501 (Giroletti et
al. 2004) is in favour of a jet intrinsic property
in the inner region (as discussed
by Meier, 2003),
whereas a jet velocity decrease
due to the ISM is likely to be present in an intermediate region (from the
pc to the kpc scale).
 
\section{Two $\it{laboratory}$ sources}

\subsection{1144+35}

1144+35 is a giant radio galaxy which shows a strong extended
jet and a short counter-jet on the pc scale (Giovannini et al., 1999).
The main jet is limb-brightened with evident substructures in the external
shear layer which are
moving with $\beta_a$ $\sim$ 2.7 (A, B, B1, and A1 in Fig. 6).
New recent observations at 8.4 GHz have allowed to measure a possible
proper motion also in the counter-jet (E in Fig. 6): $\beta_{a-cj}$ = 0.3 
$\pm$ 0.1 in the
time range 1995 - 2002 with 4 different epochs (Giovannini et al. in
preparation).
According to Mirabel and Rodriguez (1994) using both the jet
and counter jet
proper motions we can obtain directly $\theta$ = 30$^\circ$ and
$\beta$ = 0.9.
From these data we can obtain a comprehensive scenario, as follows:
1) the bulk and pattern velocity are the same;
2) the shear layer Doppler factor is $\sim$ 2 (being $\Gamma$ $\sim$ 2.3)
while assuming $\Gamma$ = 15
for the inner spine, its Doppler factor is 0.7 in agreement with
the observed brightness distribution;
3) if the external shear layer started with the same velocity of the inner
spine,
its velocity decreased from 0.998c to 0.9c in less than 100 pc suggesting
an intrinsic origin for the jet velocity structure.

\begin{figure}
\centerline{
\includegraphics[width=26pc]{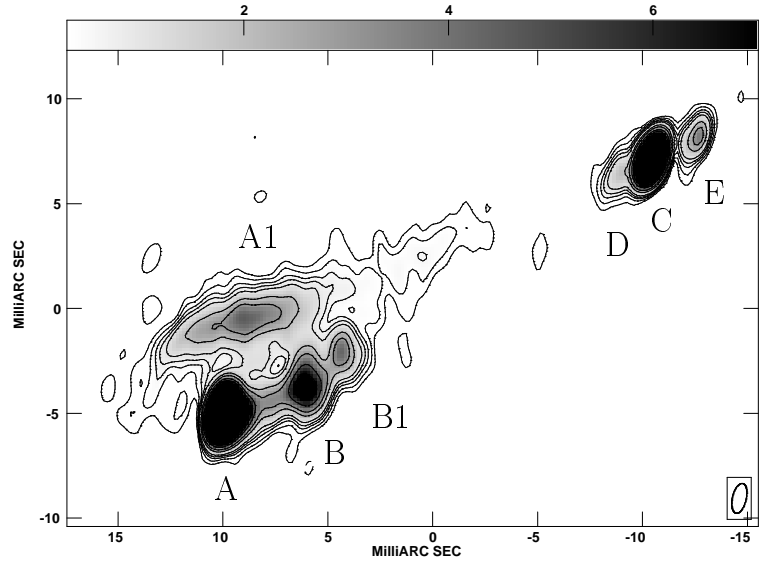}}
\caption{Global VLBI image of 1144+35 at 8.4 GHz. C is the source
core, E the counter-jet and D B1 A1 B A are substructures in the main jet.}
\label{fig6}
\end{figure}

\subsection {Markarian 501}

Markarian 501 (Mkn 501) is a nearby (z = 0.034) BL-Lac source well known
for its X- and $\gamma$- ray emission. At parsec resolution it is one  of the
best observed radio sources. In a recent paper Giroletti et al. 2004
show a clear evidence of a limb-brightened jet structure visible at a
distance $<$ 1 mas from the core up to 50-100 mas from the core.

\begin{figure}
\centerline{
\includegraphics[width=20pc]{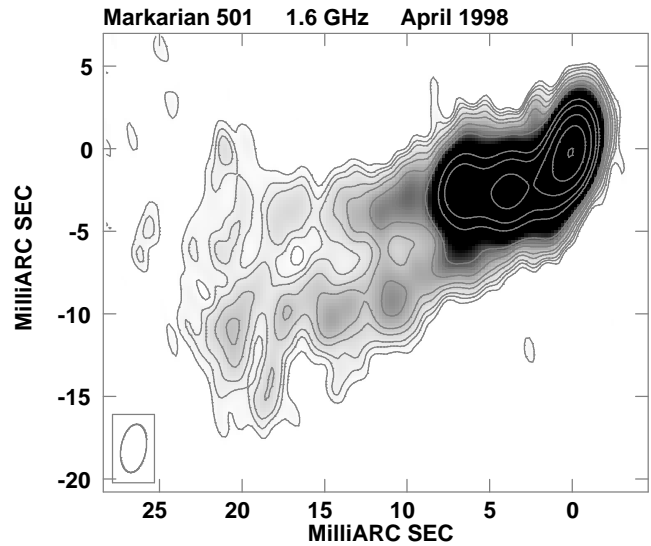}}
\caption{Space VLBI image of Mkn 501 at 1.6 GHz}
\label{fig7}
\end{figure}

The presence of a velocity jet structure very near to the nuclear emission
implies a strong deceleration if it is completely due to the jet interaction
with the surrounding medium. In agreement with Meier 2003, we suggest that a
transverse velocity gradient is intrinsic and that a further velocity decrease
due to mass loading in the jet from the ISM is likely to be dominant at larger 
distances from the core.

Giroletti et al. discuss also evidences of a highly relativistic pc scale jet 
despite of the lack
of any measured proper motion.
In this source from the high frequency ($\gamma$-ray) emission, Salvati et 
al. (1998) found that a jet with a bulk velocity $\Gamma$ $\ge$ 10 with 
opening angle and orientation angle $\sim$ 1/$\Gamma$ is required. This result
is in contrast with observational evidences of radio data that, as discussed 
in Giroletti et al., require a jet orientation angle larger than 15$^\circ$.
To reconcile the radio and high frequency results we have to assume that the
jet in the inner region ($<$ 0.03 pc) where high frequency radiation is
produced and radio emission is negligible is oriented at $\theta$ $\le$ 
5$^\circ$. Over the range from 0.03 to 50 pc the jet orientation has to
change from 5$^\circ$ to 15$^\circ$ -- 25$^\circ$. The origin of this
possible gradual change it is not yet known. We recall that beyond 30 mas
from the core the jet position angle becomes stable and closely aligned
with the kpc scale structure.

\section{Polarization}

In quasar and BL-Lacs the nuclear emission is weakly polarized at a level of 
a few percentage with
a higher fraction of polarized flux at high frequency. Radio jets have 
5 -- 10\% polarization with a tail up to a few tens of percent.
No polarized emission has been found in radio galaxies at a few percent level.

About the orientation of jet polarization with respect to the jet direction
conflicting data are reported in literature. It seems that quasar jets are
dominated by longitudinal magnetic fields but with a large tail.
BL-Lac objects have an excess of parallel oriented vectors (transverse magnetic
fields) but also in this case a large tail is present.

Present more accepted interpretation is the dominance of the longitudinal
or toroidal component of helical jet magnetic fields (Gabuzda, 2002).
We note that most parsec scale polarization observations are sensitive only
to polarization in the brightest jet regions, therefore the measured
polarization percentage and orientation could not reflect the full jet 
magnetic field properties.

In jets where a transverse structure has been found probably due to a velocity
structure (sect. 5), a parallel magnetic field is present in the sheat boundary
layer and a perpendicular magnetc field in the inner spine.

\section{From pc to kpc}

Radio jets in low power (FR I) radio galaxies are characterized on the kpc
scale by a symmetric two-sided structure, a large opening angle,
and the presence of a magnetic field perpendicular to the jet axis.
Observational data show the presence of a strong jet deceleration within
$\sim$ 5 kpc from the core (see e.g. 3C449, Feretti et al. 1999). 
Large scale jets
are therefore low velocity jets and often show morphological distortions
as oscillations or large curvatures (head-tail or wide-angle-tail radio
galaxies) due to a strong interaction with the ambient medium.

Jets in high power (FR II) radio galaxies and quasars are very collimated,
in many cases strongly asymmetric (one-sided), and with a magnetic field 
parallel to the jet axis.
Relativistic beaming affects the observational properties of pc and kpc
scale jets, however peripheral regions are less dominated by Doppler boosting,
suggesting a decreasing of the jet Lorentz factor from the pc to the kpc scale.
A value of $\Gamma$ $\sim$ 2 -- 1.5 is consistent with kpc scale 
properties in many
large scale FR II sources (see also Arshakian and Longair, 2003).

\section{Conclusions}

In recent years we have seen a large improvement in number and quality of
observational data of jets. From these data we can derive the following
conclusions:

\begin{itemize}

\item
The parsec scale jet velocity is highly relativistic with $\Gamma$ in the range
3 to 10 in high and low power radio sources. The jet velocity is not correlated
to the total or core radio power.

\item
The jet morphology in the pc scale is the same in high and low power radio 
sources.

\item
In some sources the pattern and bulk velocity are the same.

\item
A two velocity regime is necessary to explain some observational properties
in low power sources and possibly in a few high power sources. A different
Doppler factor can explain the limb-brightened structure observed in some 
sources. 

\item
It is not yet clear if the jet velocity structure is a jet intrinsic property 
or it is due to the jet interaction with the ISM. In any case at some distance
from the core the jets slow down because of interaction with the ISM.

\item
BL-Lac sources show an excess of magnetic field perpendicular to the jet
direction while Quasars show an excess of longiudinal magnetic fields, 
however a broad tail is present.

\item
In jets with a velocity structure, the boundary layer show a parallel magnetic 
field with respect to the jet direction, 
and the inner spine a perpendicular magnetic field. Radio galaxies are not
polarized at a few percent level.

\item
Large scale jets in FR I have a low velocity and are not relativistic. 
They decelerate drammatically from the pc to the kpc scale. 
In FR II sources jets decelerate more slowly. They can move with a Lorentz 
Factor $\sim$ 2 also at large distance
(kpc scale) from the core.

\end{itemize}

\acknowledgements
The author thanks the Scientific and Local organizing people for the
wonderful idea to organize such a good meeting.
This work was partially supported by the Italian Ministry for University and 
Research (MIUR) under grant COFIN2001-02-8773.

\end{article}
\end{document}